\documentclass{article}
\usepackage{spconf,amsmath,graphicx,subcaption,ulem}
\usepackage{xcolor}
\newcommand{\Tau}{\mathrm{T}}

\title{Towards an intelligent microscope: adaptively learned illumination for optimal sample classification }
\name{Amey Chaware*, Colin L. Cooke*, Kanghyun Kim, Roarke Horstmeyer }
\address{Pratt School of Engineering, Duke University, Durham, NC 27708, USA \thanks{*Equal Contribution}\thanks{Data and code: deepimaging.io/recurrent-illuminated-attention}}

\begin{document}
%
\maketitle{}
\begin{abstract}
Recent machine learning techniques have dramatically chan-ged how we process digital images. However, the way in which we capture images is still largely driven by human intuition and experience. This restriction is in part due to the many available degrees of freedom that alter the image acquisition process (lens focus, exposure, filtering, etc). Here we focus on one such degree of freedom - illumination within a microscope - which can drastically alter information captured by the image sensor. We present a reinforcement learning system that adaptively explores optimal patterns to illuminate specimens for immediate classification. The agent uses a recurrent latent space to encode a large set of variably-illuminated samples and illumination patterns. We train our agent using a reward that balances classification confidence with image acquisition cost. By synthesizing knowledge over multiple snapshots, the agent can classify on the basis of all previous images with higher accuracy than from naively illuminated images, thus demonstrating a smarter way to physically capture task-specific information. 
\end{abstract}

\section{Introduction}
\label{sec:intro}

\begin{figure}[tb]
    \centering
    \includegraphics[width=0.38\textwidth]{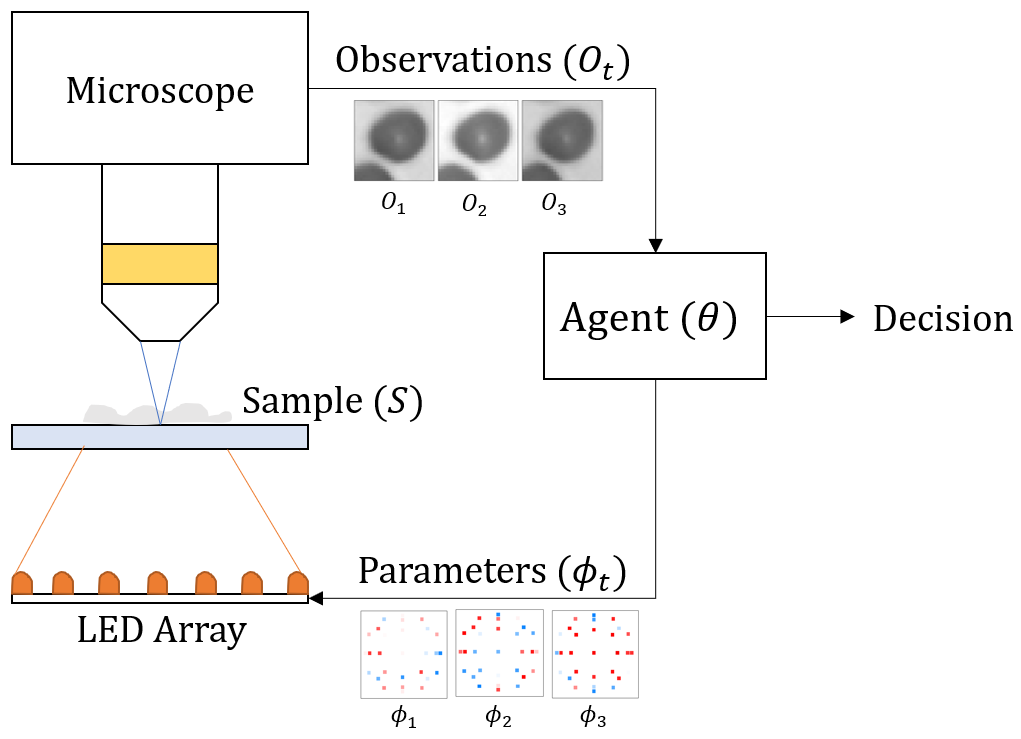}
    \caption{Our intelligent microscope uses a reinforcement learning agent, parameterized with $\theta$, to control an LED array ($\phi$) across multiple timesteps $T = \{1...T\}$ to make a decision.}
    \label{fig:agent_env}
\end{figure}

The optical microscope remains a critical tool in medicine and biology. Over the past several decades, advances with digital microscopes and associated post-processing algorithms have led to fundamental improvements in resolution \cite{hell2007far}, 3D imaging depths \cite{ouzounov2017vivo} and acquisition speed \cite{gao2014single} - opening up new avenues for scientific discovery. At the same time, these improvements have led to increasingly more complex microscope devices and astronomically larger datasets.

To help us handle the increasingly large quantities of raw data that we detect about various biological systems, many are now turning to machine learning. However, the majority of machine learning methods are currently used to process ``standard" microscope image data that has already been digitally captured and saved. These algorithms do not influence how data is captured and make no effort to improve the quality or relevance of the images.

In this work, we aim to change this paradigm, by creating a sensing framework for an “intelligent” microscope, whose physical parameters (e.g., illumination settings, lens configuration, and sensor properties) are adaptively tuned to capture better task-specific images that contain the most relevant information for each learning goal. To achieve this, we integrate within a deep learning post-processing system an attention mechanism, which models active control of the microscope. Using closed-loop feedback and reinforcement learning, we optimize for a series of measurements, captured under different experimental microscope settings, to maximize the performance of our automated decision software.

Effectively, this principle affords the microscope a large amount of flexibility to extract useful information from the sample of interest. Much in the same way as we physically interact with a new object when trying to understand what it is, we aim here to allow the microscope to reason through and execute physical interactions to form decisions about each sample with as high as accuracy as possible. Clearly, this approach is directly related to reinforcement learning techniques \cite{lillicrap2015continuous}, in which agents iteratively explore and interact with their environment to complete a task.

Our first aim for this new method to help with the general problem of improving the quality of measurements for automated microscope image diagnosis. Almost always, samples extracted from patients are either too large to fit within the microscope’s field-of-view, or too thick and thus significantly scatter light. Within this paper we will examine two classification problems. The identification of the malaria parasite within experimentally collected images of blood smears, and a heavily modified version of the MNIST digit recognition problem. Although these tasks are both classification-based, we argue that the presented paradigm can be generally applied to any high-dimensional, controllable, measurement scheme with a single desired artificial intelligence-driven outcome.
\section{Previous Work}
\label{sec:relarted work}
There are several recent works that consider use of machine learning to jointly optimize hardware and software for imaging tasks \cite{Sitz2018, Kellman19,Diederich18, ayan2016,Barbastathis19,Chang18,Hershko19}. These approaches aim to find a fixed set of optical parameters that are optimal for a particular task. Although these methods provide improved performance relative to standard parameterizations, their results are only optimal for a single task, and do not provide any means of adaptability across individual samples.

Two recent works \cite{roarke17,optimized_segmentation} have studied the impact of optimizing a programmable LED array over entire datasets. These works have shown that a fixed optimal pattern yields increased performance in both classification and image-to-image mapping tasks. These works show that inclusion of programmable illumination within a microscope allows joint hardware-software optimization using machine learning. The programmable hardware leads to better task-specific performance and can be tuned without requiring any moving parts.

Adaptively choosing imaging parameters is a relatively underexplored area. Some works like Yang et al. \cite{rlexp18} consider the use of reinforcement learning for real time metering and exposure control. However, no work has yet aimed to dynamically change acquisition parameters during multi-image capture in response to the contents of the sample.

Visual attention mechanisms allow machine learning algorithms to self-select visual information which is most relevant for further processing. The concept of \textit{recurrent visual attention} was first shown in Mnih et al. \cite{ram14} where a recurrent policy learned to iteratively sample a ``glimpse" of the x-y plane. Further work has both reinforced the performance of recurrent visual attention \cite{ram2015} and expanded the attention mechanism \cite{xu2015show}. 
\section{Methods}
 
\begin{figure}[tb]

\begin{minipage}[b]{1.0\linewidth}
  \centering
  \centerline{\includegraphics[width=7.5cm]{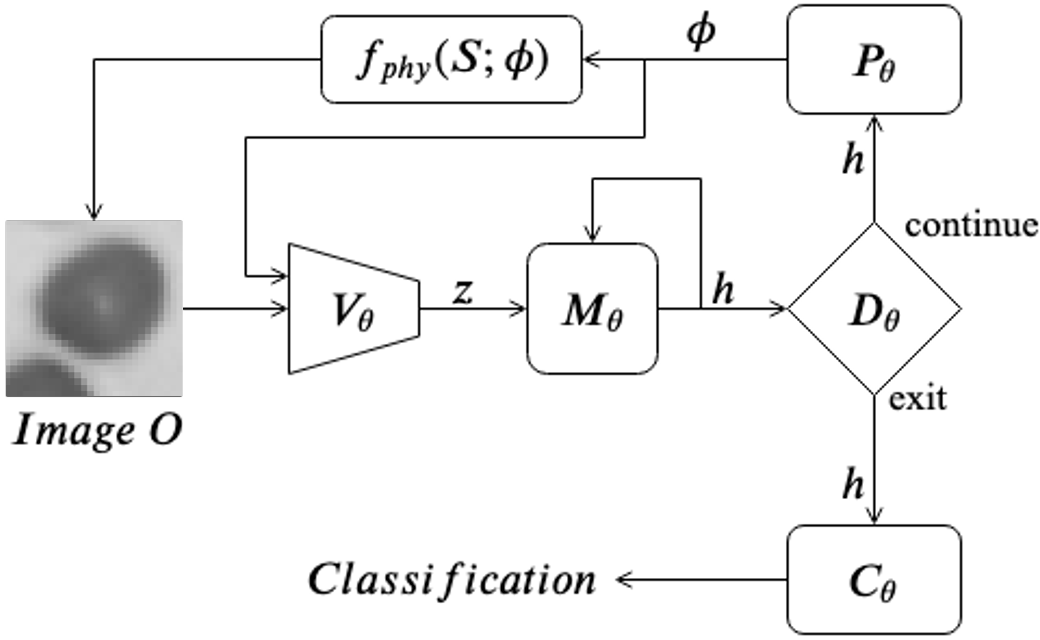}}
\end{minipage}
\caption{
Flow diagram of our model: $V_{\theta}$ denotes the Visual Encoder, which creates an embedded representation, $z$, of the image $O$. $h$ is the hidden state of the Memory unit $M_\theta$. $P_{\theta}$ is the Parameter Network, which samples the parameters for the next iteration. $D_{\theta}$ is the Decision Network, which decides if enough information has been gathered to perform the classification, which the Classification Network, $C_{\theta} $ performs.}
\label{fig:network}
\end{figure}
\subsection{Adaptive Sensing}
We consider an agent interacting with a visual environment (i.e., the sample of interest) via an imaging system. The agent has direct control of the parameters of the imaging system through a visual attention mechanism, allowing the self-selection of information required to accomplish its task. The system is bandwidth limited, since the interaction of light with both the lens and sensor prevent complete capture of environment state within a single image, in response the agent is required to synthesize information across multiple time steps to build an accurate state representation. At each step, the agent is given a choice, to either make a decision about the sample using the information gathered so far (e.g., predict its class), or to select new information through the integrated attention mechanism (i.e., capture a new image under different illumination). This choice presents a trade-off between acquisition cost and task performance, which can be tuned given the needs of overarching system.
\begin{figure}[tb]
    \centering
    \includegraphics[width=7.5cm]{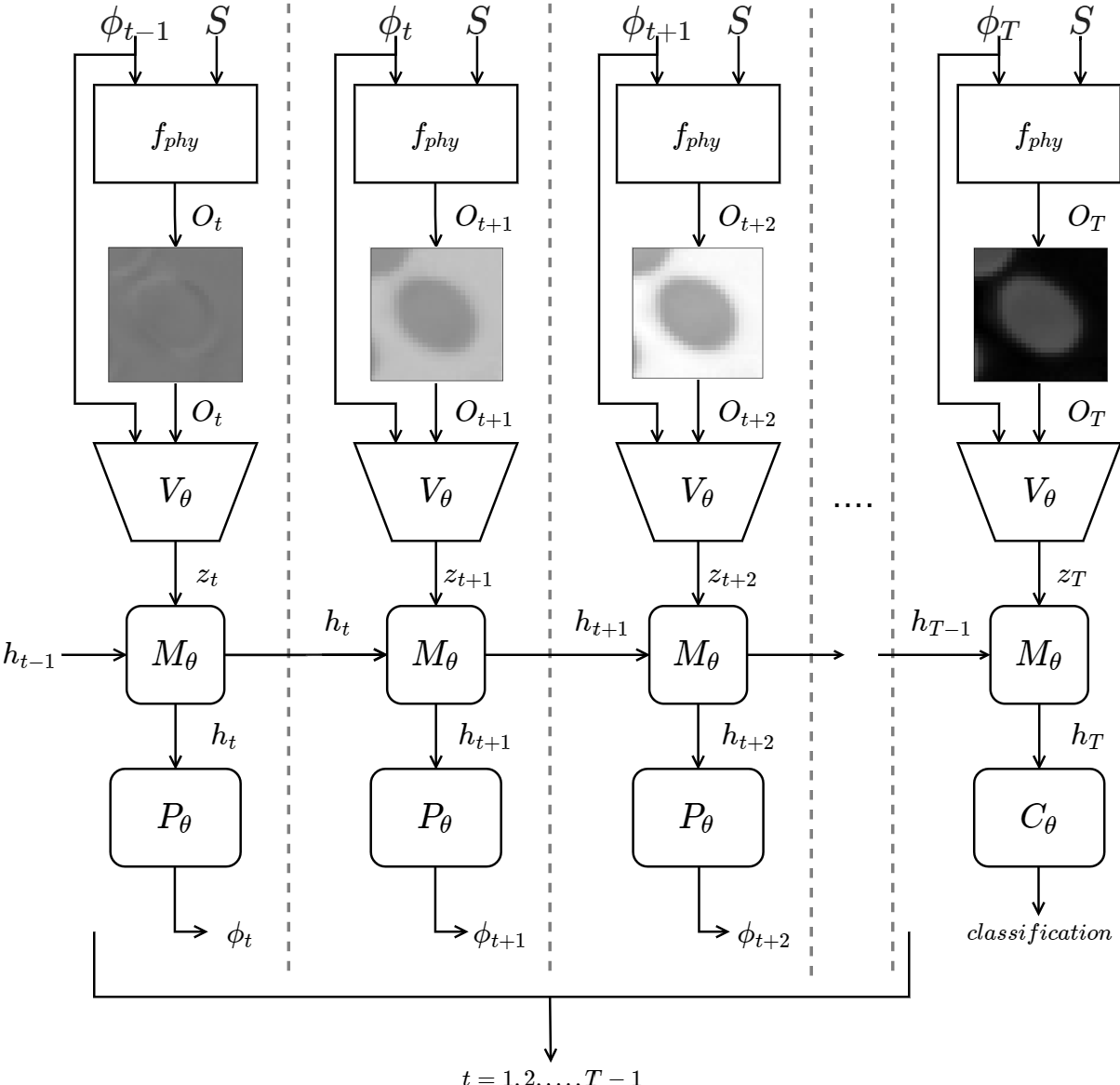}
    \caption{Our agent applied to a $\Tau$ length trajectory of observations. The hidden state of the memory element $M$ is used to pass information between steps, the physical parameterization $\phi_{t+1}$ is drawn from $P_\theta$ and conditioned on $h_t$. At $t=\Tau$ a classification is made using the final hidden state $h_\Tau$}
    \label{fig:unrolled}
\end{figure}
The image formation process is modelled as $f_{phy}$, with the formed image from a system parameterized by $\phi$ of a given sample $S$ given by: $O = f_{phy}(S;\phi)$. An agent (Figures \ref{fig:network} and \ref{fig:unrolled}) is constructed which optimizes the configuration of the hardware across multiple acquisitions jointly with the post-processing of images. A Visual Encoder, $V_\theta$, is used to convert each observed image into an embedded representation: $z_t = V_\theta(f_{phy}(S;\phi_{t-1}),\phi_{t-1})$. This representation is fed into a recurrent unit, $M_\theta$, which maintains the agent's internal state by synthesizing embeddings across the history of observations: $h_{t} = M_{\theta}(z_t, h_{t-1})$.

At each step, the agent's decision engine ($D_\theta$) decides if enough information has been gathered to make an accurate classification, or if more samples are required. Depending on this outcome a new system parameterization $\phi$ is produced by the Parameter Network $P_\theta$ or a classification decision $\hat{y}$ is made using the Classification Network $C_\theta$:
\begin{align*}
d_t &\sim D_{\theta}(d_t | h_t)\in \{1,2\}\\
 & \begin{cases}
            \phi_t \sim P_\theta(\phi_t | h_t) & \text{if } d_t = 1\\
            \hat{y} = C_{\theta}(h_t) & \text{otherwise}\\
        \end{cases}
\end{align*}
The implemented agent uses two layer fully-connected networks for $V_\theta$ and $D_\theta$, while $C_\theta$ and $M_\theta$ use a single fully-connected layer and a 256 unit LSTM respectively.

\subsection{LED array illumination}
In our work, we consider the microscope as the visual environment, and optimize over the crucial element of illumination.
Using a programmable LED array, we can design an optimal light pattern, which could be a mixture of bright-field and dark-field illumination, to highlight sample features that are important for a particular task \cite{Muthumbi:19,optimized_segmentation}.

Consider the image $I_k$ of the sample $S$ formed by turning on the $k$th LED in the array with brightness $w_k$. It can be written as, $I_k = f_{phy}(S;w_k)$, where $f_{phy}$ is the physical model of image formation.
Under linear image formation, $I_k$ can be found by multiplying $w_k$ with the formed with the $k$th LED at a fixed brightness, $\hat{I}_k$
\[
I_k = f_{phy}(S;w_k) = w_k f_{phy}(S;w_k=1)=w_k \hat{I}_k.
\]
In absence of noise and assuming each LED is mutually incoherent, the image formed by turning on multiple LEDs is equal to the computed sum of the images captured from each LED turned on individually~\cite{roarke17}. Specifically, if the brightness of the $k^{th}$ LED is denoted as $w_k$ and the associated image formed by illuminating the sample \emph{with this LED only} at a fixed brightness is $\hat{I}_k$, then the image $O$, formed by turning on a set of $N$ LEDs at variable brightness is given by:
\begin{equation*}\label{eq:image_formation}
    O = \sum_{k=1}^N w_k\cdot f_{phy}(S;w_k=1) = \sum_{k=1}^N w_k \hat{I}_k
\end{equation*}
The set of LED brightnesses $\{w_k^{(t)}\}_{k=1}^{N}$ is denoted as $\phi$, which is the parameterization of the imaging system such that $O = f_{phy}(S;\phi)$. We use this image formation process to construct a visual attention mechanism which uses LED brightnesses ($\phi$) select information from the underlying sample $S$.
\subsection{Data preparation}
For the MNIST based task we simulate microscope image formation under variable illumination using the MNIST datset. Each normalized MNIST image was used to define the height profile of thin, translucent sample (maximum  thickness $t=2.5\ \mu m$). The profile was then sequentially illuminated by 25 distinct LEDs placed in a $5\times5$ grid located 50mm beneath the sample at a 6mm pitch. The optical fields of each LED were propagated through the sample and a simulated objective lens with $5\times$ magnification and 0.175 NA. Finally, a simulated detector with $28\times28$, $7\mu$m pixels received the optical field to form an image, exhibiting Gaussian noise in readout. The final training set consisted of 60000 samples while the test set contained 10000. Each image set was stored as a $28\times28\times25$ tensor. 

Our experimental dataset consists of images of blood cells from thin smear slides used in the clinic to diagnose malaria (experimental details in Ref.~\cite{roarke17}). The images were cropped and labelled to construct a binary classification task of diagnosing the presence of the malaria parasite. Variable illumination was provided by 29 LEDs, where each LED contained 3 individually addressable spectral channels, creating 87 uniquely illuminated images. The images were augmented by flipping and rotation, creating a total of 4100 samples stored as $28\times28\times87$ tensors. Train and test sets were constructed with an 80-20 split.



\begin{figure}[t!]
    \centering
    \begin{subfigure}[t]{0.35\textwidth}
        \centering
        \includegraphics[width=\textwidth]{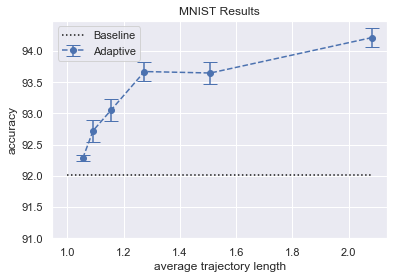}
        \subcaption{MNIST results}
        \label{fig:MNIST_results}
    \end{subfigure}%
    \\
    \begin{subfigure}[t]{0.35\textwidth}
        \centering
        \includegraphics[width=\textwidth]{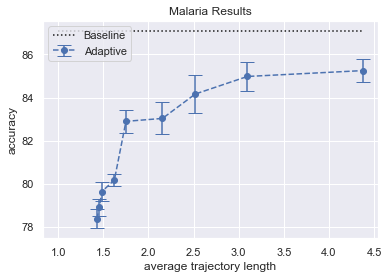}
        \subcaption{Malaria results}
        \label{fig:malaria_results}
    \end{subfigure}
    \caption{Performance acquisition curves}
    \label{fig:results}
\end{figure}

\subsection{Training}
Our adaptive machine learning model uses two distinct outputs to accomplish its task. The first is a classifier which translates the hidden state into a classification decision. The second is a decision engine, which evaluates overall classification confidence, and decides if enough information has been captured to make a correct classification, thus ''exiting" the feedback loop. The exit mechanism provides a means to evaluate samples over variable length trajectories, capturing more data when necessary and exiting to perform a decision when appropriate.

A successful exit is then defined by an exit followed by a correct classification, with an unsuccessful exit being the opposite. We assign a reward to each of these outcomes, allowing control of the trade-off between acquisition cost and accuracy. The decision engine is trained using this outcome dependent reward, mapped to a cross entropy loss function, and the classifier uses categorical cross-entropy based loss. While the classification loss doesn't affect the decision model and vice versa, both of the losses influence the other portions of the model.


\section{Results}
\label{sec:experiments}
\begin{figure}[t!]
    \centering
        \centerline{\includegraphics[width=0.40\textwidth]{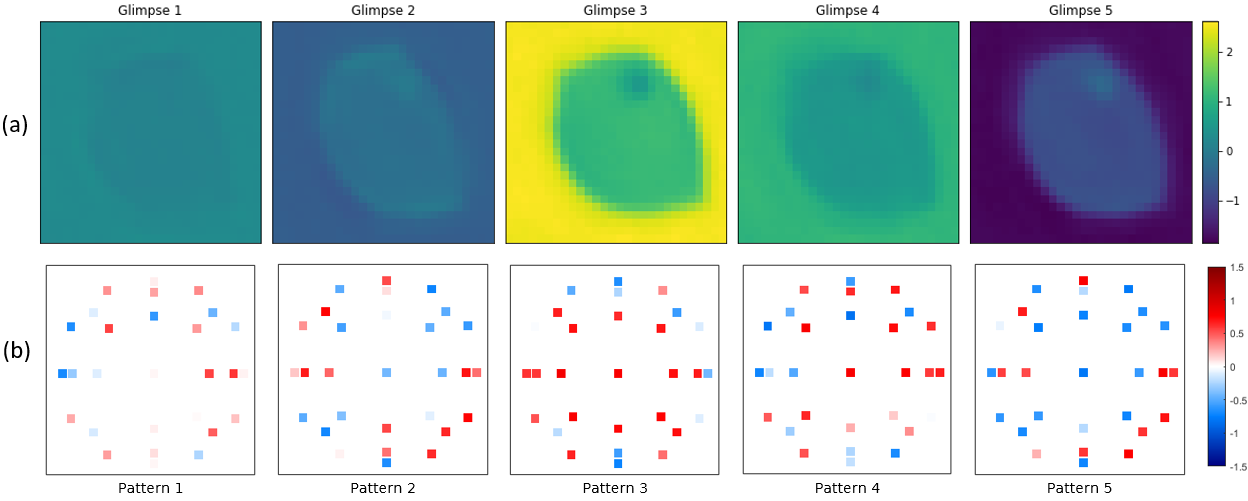}}
     \caption{(a) shows progression of observations obtained by simulating the image formation process and (b) shows the corresponding LED arrays generated by the agent. For brevity only the green channel of the LED patterns is shown.}
     \label{fig:sequence}
\end{figure}

We conducted two experiments to understand the benefit of an adaptive attention mechanism using the simulated MNIST and  physically collected malaria image data. For each experiment we ran both a baseline (a single forward pass of the agent's model, with a fixed learned LED pattern) and our new adaptive approach. Using the same forward model in each case allowed a fair comparison between the approaches, uninfluenced by network architecture. We evaluated the agent's performance across a sweep of exit decision rewards ($R_{exit}$) from $\{1.0 \rightarrow 0.01\}$ while the stay decision reward ($R_{stay}$) remained fixed at $1.0$, which allowed us to examine performance as a function of image capture number. Multiple random seeds are used for each configuration to obtain variance estimates. Fig \ref{fig:sequence} shows a visualisation of the trajectories that the network took for a sample, where we observe that the agent prefers to probe the samples under a variety of different illumination schemes during the decision process.
These results show that an adaptive sampling paradigm can effectively trade-off task performance for acquisition cost. Both the MNIST and Malaria datasets (Figure \ref{fig:results}) show a decaying positive relationship between average trajectory length and overall system accuracy.
Additional information drives this trajectory length to accuracy relationship - longer trajectories allow the agent to gather more information about the sample prior to making a decision. 
We observe that more information isn't always better, however - some systems aren't limited by the information that is available, but instead by their ability to process it. We hypothesize that this is the case with the malaria classification task, where although the positive relationship does exist, a more direct training paradigm (the baseline method) offers higher performance.


\section{Conclusion}
This work establishes a reinforcement learning based framework to deploy an adaptive imaging system, where an attention mechanism is used to perform data-dependent sampling. Although our experiments are exclusive to controlling illumination for a classification task, we expect our framework to extend to more elements of the microscope and different kinds of tasks. By including an exit mechanism within the recurrent structure, we establish a relationship between trajectory length and task performance. We postulate that this relationship is not only task dependent, but also depends on the processing capability of the network. A classification system may not be fundamentally information limited, however, as the amount of information demanded from the task increases (such as in an image-to-image inference task), we expect this trajectory-to-performance curve to shift. We demonstrate that adaptive systems not only offers increased performance compared to fixed imaging systems, but also show a path towards the integration of data capture and processing.
\newpage
\bibliographystyle{IEEEbib}
\bibliography{refs}

\end{document}